\documentstyle[12pt]{article}

\begin{document}

\title{Statistical Mechanics and Quantum Cosmology
\thanks{
Invited talk given at the Second International Workshop on Thermal Field
Theories and Their Applications, Tuskuba, Japan, July 1990.
Published in {\it Thermal Field Theories}, edited by H. Ezawa, T. Aritmitsu
and Y. Hashimoto (North-Holland, Amsterdam, 1991) p. 233-252.}}
\author{B. L. Hu
\thanks{Work supported in part by the National Science Foundation under
Grant No. PHY87-17155.}\\
{\small Department of Physics, University of Maryland,
College Park, MD 20742, USA }}
\date{\small {\sl (UMDPP\#91-096, Nov. 1990)}}
\maketitle
\begin{abstract}
Statistical  mechanical  concepts  and  processes  such  as  decoherence,
correlation,  and  dissipation  can  prove  to  be  of  basic  importance  to
understanding some fundamental issues of quantum  cosmology  and  theoretical
physics such as the choice of initial states, quantum to classical transition
and the emergence of time.  Here we summarize our effort in 1) constructing a
unified theoretical framework using techniques in interacting  quantum  field
theory such as influence functional and coarse-grained  effective  action  to
discuss the interplay of noise, fluctuation, dissipation and decoherence; and
2) illustrating how these concepts when  applied  to  quantum  cosmology  can
alter the conventional views on some basic issues.  Two questions we  address
are 1) the validity of minisuperspace truncation, which  is  usually  assumed
without proof in most discussions, and 2) the relevance of  specific  initial
conditions, which is the prevailing view of the past decade.  We also mention
how some current ideas in chaotic dynamics, dissipative collective  dynamics
and complexity can alter our view of the quantum nature of the universe.

\end{abstract}

\newpage
\section{Introduction}

In this talk I would like to present some general personal views  on  how
the concepts and methodologies in statistical mechanics  can  be  of  use  to
facilitate a better understanding and a clearer  formulation  of  some  basic
issues of quantum cosmology.
\par
In keeping with the interdisciplinary nature of this workshop I will only
discuss ideas here and avoid technicalities, knowing that one can always find
the details in the  original  papers.   Because  of  the  general  nature  of
discussions the viewpoints presented here could also appear to be  tentative,
sketchy and partial.  I will, however, try to mention the  important  papers,
as well as recent reviews and conference proceedings in this field for  those
who would like to get a better overall perspective.  In this connection I may
suggest Proceedings of the 1988 Osgood Hill Conference$^{1}$, the  1989
Jerusalem Winter School$^{2}$, and selected papers in the 1989 Santa Fe
Institute  Studies$^{3}$. Quantum Cosmology began in the sixties with the work
of DeWitt$^{4}$ Wheeler$^{5}$  and Misner$^{6}$.
The revival of interest in the eighties was brought about primarily
by the works of Hartle, Hawking$^{7}$ and Vilenkin$^{8}$. A
very complete bibliography of recent papers on quantum cosmology was
compiled by Halliwell$^{9}$, who himself contributed extensively to its
recent development.
\par
Information theory  and  quantum  measurement  ideas  useful  in  quantum
cosmology were introduced by  Wheeler,  Zurek  and  Unruh$^{10}$,  Joos,  Zeh
and Kiefer$^{11}$,  Griffith  and  Omnes$^{12}$,  Gell-Mann  and
Hartle$^{13}$.    Since   this
particular aspect of statistical mechanics in relevance to quantum  cosmology
is quite well noticed (see Ref. 3), I will not pursue it here. Neither will I
belabor on the Hamiltonian$^{4-6}$ nor the path  integral$^{7,8}$
formulations,  which
make up most of the formal work in this field.   Because  I  have  difficulty
understanding  the  physical  meaning  of  results  derived  from   Euclidean
techniques I prefer to start from a firmer ground (quantum  field  theory  in
curved  spacetime,  or  quantum  mechanics)  and  analyze  smaller   problems
(Brownian motion  and  the  ubiquitous  harmonic  oscillator!)  which  I  can
understand and trust the results.  In this sense I'll actually not  be  doing
quantum cosmology (as we shall  see  these  two  words  given  their  present
connotation  can  be  intrinsically  contradictory),  but   discuss   quantum
statistical  processes  like  dissipation,   fluctuation,   correlation   and
decoherence in a much simpler context.  These problems, which  are  basic  to
many physical phenomena and are believed to play an important role in quantum
cosmology,  have  simply  not  been  understood  well  enough  in  the   more
complicated conditions characteristic of quantum cosmology to warrant liberal
generalization and excessive claims.  In  essence,  then,  I  will  take  the
layman's approach, asking  some  intuitive  and  rudimentary  questions,  and
trying to see if the basic tenets of quantum  cosmology  are  sound  and  are
compatible with what we understood in statistical and quantum  physics.   One
of these sample  problems,  specifically  on  the  stochastic  properties  of
interacting quantum fields, has recently been treated in  detail  from  first
principles by Juan Pablo Paz, Yuhong Zhang and myself$^{14-17}$.  It was
reported in Dr. Paz's talk.  We are now in the process of  trying  to
understand  its implications  in  the  context  of  stochastic   inflationary
universe$^{18-22}$ semiclassical gravity$^{23.24}$, and quantum cosmology.
The study of  dissipation in quantum fields, semiclassical gravity and in
quantum  cosmology  has  been pursued by Esteban Calzetta and myself over the
years$^{25-33}$ and  furthered  by Paz$^{34,35}$ recently.  An earlier account
of some views on the basic  issues  of
minisuperspace cosmology can be found in my 1989 Erice lectures$^{31}$.

\section{Viewpoint and Issues in Quantum Cosmology}

Classical cosmology as a discipline, from Newtonian to  relativistic,  is
based on the continuum notion of spacetime described by Riemannian  geometry.
There is no valid picture of  quantum  geometry  as  yet.   When  gravity  is
conceived as the force mediated by a spin-two particle, the graviton, one can
talk about quantum gravity, even though grave problems exist in  relation  to
its renormalizability; and for  its  higher  derivative  extensions,  the
$R^{2}$
theories, also  unitarity  and  causality  problems.   Einstein's  relativity
theory based on the Hilbert action  $\int R \sqrt {-g} d^{4}x$
from  which  cosmology  is
derived, is in this view believed to be the long  wavelength  or  low  energy
limit of a more complete and intricate theory, superstring theory  being  one
serious candidate.  Above the Planck scale quantum  properties  of  spacetime
become  important.   Here   geometries   with   non-trivial   topology   can
spontaneously be created and  annihilated,  rendering  the  smooth  continuum
picture of spacetime  totally  inadequate.   At  this  time  there  are  many
attempts at constructing a picture of quantum spacetime but no success is yet
in sight.  The goal is however shared by workers on  many  fronts,  including
superstring theory$^{36}$, conformal field  theory$^{37}$,topological  field
theory$^{38}$, wormholes and baby universes$^{2}$, and random geometry$^{39}$.
\par
Therefore, according  to  this  view,  any  talk  on,  say,   superstring
cosmology, which starts with a Robertson-Walker or de Sitter metric does  not
make much sense, because the spacetime constructed  from  and  governing  the
superstrings cannot be the simple manifolds.  One can at best be  looking  at
the low energy limit (about, but not  above  the  Planck  scale)  of  such  a
theory, which can equally well be treated by  the  well-known  semi-classical
theories.  If one  views  gravity  as  an  effective  theory  resulting  from
elementary particle interactions, as Sakharov$^{40}$ had proposed, the  notion
of
continuum spacetime only  makes  sense  in  the  long-range  limit,  much  as
elasticity is to electrodynamics.  Then quantum cosmology is as meaningful as
quantum elasticity.  It is almost just as crazy (though doable in  principle)
to try to derive elasticity from QED, as is it hopeless (though we  all  seem
to be trying) to deduce QED from elasticity.  The incongruence  ingrained  in
the words "quantum cosmology" is a reflection of the deeper dilemma one faces
today in probing the quantum structure of spacetime.
\par
So just  what  exactly  does  one  mean  when  one  talks  about  quantum
cosmology? and what does one gain in asking such almost impossible questions?
Different people have  different  ideas  about  quantum  cosmology.   If  one
believes that the metric  is  a  basic  observable  depicting  a  fundamental
physical field quite at the opposite spectrum from the effective theory, then
one would quantize the three-geometry $^{3}g_{ij}$ by imposing commutation
relations between $^{3}g_{ij}$ and $\pi _{ij}$, its conjugate momentum,
like what one usually does with
any classical physical observable.  The difficulties one encounters are  that
of quantum gravity proper.  Quantum cosmology then refers to the quantization
of a restricted set of the degrees of freedom  allowed  in  quantum  gravity.
The  geometries  commonly   encountered   in   (classical)   cosmology   like
Robertson-Walker, de Sitter, and mixmaster universes are only the homogeneous
anisotropic solutions to Einstein's equations.  In a perturbative sense (e.g.
the Lifshitz operator) one can view these geometries as the lowest  modes  of
spacetime excitations$^{31}$, the higher modes corresponding to the
inhomogeneous universes are being ignored.
In the superspace picture$^{4,5} ($the space of  all 3-geometries) homogeneous
cosmology constitutes a highly truncated class, the so-called minisuperspace
$^{6}$.  The problem is now simplified from quantizing  an
infinite to only a few degrees  of  freedom.   Most  discussions  of  quantum
cosmology make such a simplification.  Whether this is acceptable has been an
open question since the Hamiltonian  cosmology  of  the  sixties.   We  shall
address this issue concerning the validity of minisuperspace  approximation
$^{41}$ from the statistical mechanical viewpoint later.
\par
Quantum cosmology in the eighties works also mainly with  minisuperspace,
i.e., makes the same assumption of truncated  degrees  of  freedom,  but  the
attention was shifted to the issue of initial states or  boundary  conditions
on the Euclidean path integrals.  This opened up new avenues because a simple
prescription$^{7,8}$ on what 4-geometries  to  sum  over  in  the  Euclidean
path
integral leads to physically interesting results.  For  example,  Hartle  and
Hawking's$^{7}$ no boundary condition of summing  over  all  compact
4-geometries with boundary on $^{3}g$ and Vilenkin's$^{8}$ choice of outgoing
modes  on  regions  of
the boundary where  the  4-geometry  is  singular  both  have  the  following
desirable features:
\par
\noindent (1)  There are  regions  where  the  wave  function of the universe
$\Psi $  is oscillatory, corresponding to classically allowed solutions.
\par
\noindent (2)  There exist solutions with inflationary behavior (see, however,
Ref. 2)
\par
\noindent (3)  These boundary conditions on $\Psi $  select  a  particular
solution  to  the
functional Schr\"odinger equation and define a particular vacuum state  in  the
semiclassical limit of quantum field theory in curved spacetime.
\par
Simply put, these boundary conditions  admit  classical  solutions,  some
induce inflation, and give the correct  vacuum  state  in  the  semiclassical
limit.  The prediction of classical spacetimes corresponding to the existence
of oscillatory wave  function  is  a  very  important  result,  because  many
features of our physical world$^{18.19}$ are connected with  the  existence
of  a late universe (the flatness, age and entropy problems in  cosmology
are  all related to this fact$)^{20}$ described by classical spacetimes.
Time  is  one  of
such features, as  it  is  the  only  observable  in  quantum  mechanics  not
represented by an operator, but enters  as  a  parameter,  thus  lacking  the
interference effect intrinsic in  all  quantum  phenomena.   Because  of  its
preferred status, time plays a special  role  in  quantum  mechanics  and  in
general relativity, and brings about special problems in quantizing  gravity.
In this view, the issue of time is naturally linked to the issue  of  quantum
to  classical  transition.   In  addressing  these  two  issues   statistical
mechanical considerations enter in a  fundamental  way.   Though  occuping  a
central position, they  are  not,  however,  exclusive  concerns  of  quantum
cosmology. In fact the basic mechanisms which  can  bring  about  quantum  to
classical transitions like decoherence and  correlation  are  common  to  all
quantum phenomena and, in my view, should be better explored first outside of
quantum cosmology without its particular problems.  What makes  these  issues
particularly relevant in quantum cosmology are 1) the appearance of classical
spacetime and the associated special features of time in  classical  physics,
and 2) the existence of a late classical universe as "a  consequence  of  the
specific condition in a more general sum-over-history  framework  of  quantum
prediction"$^{13}$.  Thus according to this view$^{7,13}$  quantum  cosmology
provides
one with a pathway to connect these issues (of time and classicality) back to
the issue of initial conditions.  A  parallel  development  also  claims  the
usefulness of these initial conditions to  predict  the  value  of  universal
coupling constants in the context  of  summing  over  non-trivial  topologies
(wormholes and baby universes$)^{2}$.  The overwhelming emphasis in recent
work on
quantum cosmology is on the specificity of initial conditions.  This is where
our view differs:   Put succinctly,  we  attach  equal  importance  to  these
issues (time and classicality) which are amplified in the context of  quantum
cosmology, but we don't see the choice of  initial  conditions  as  the  most
natural way to resolve these issues.  Rather, we view the emergence  of  time
and the quantum to classical transition  more as consequences of dynamics and
interactions of the system of interest with its  environment,  and  attribute
more importance to the working  of  statistical  mechanical  effects  in  the
broadest senses, including the guiding principles of  information-theory  and
chaotic dynamics.  Let me explain what we mean by this.

\section{How are Statistical Considerations Relevant?}

In a broad sense, statistical mechanics deals with the issue  of  how  to
extract from a large, often infinite, degrees of  freedom,  a  few  variables
which can capture the most essential physics of the whole  system.   Examples
in physics are abound:  Thermodynamics describes a system in equilibrium with
its surrounding by a few macrovariables like temperature, pressure,  entropy,
etc; hydrodynamics with the transport functions, and critical phenomena  with
the critical exponents and universality classes.   In  all  cases  the  skill
rests upon 1) identifying (separation) the relevant variables describing  the
system of interest (often this involves not  just  taking  a  subset  of  the
primitive microvariables but taking the average  in  some  approximation  and
working  in  a  different  level  of  structure),  2)  averaging  away   some
information of the irrelevant variables (coarse-graining) which make  up  the
environment, and 3)  consideration  of  its  overall  effect  on  the  system
(backreaction).  This schema of  isolating  a  system  of  interest  and  the
effective accounting  for  its  interaction  with  the  surroundings  is  not
exclusive to statistical mechanics.   In  field  theory  renormalization  and
coupling hierarchy problems share the same spirit$^{29}$.
\par
Such a reduction procedure applied to the overall system and  environment
(closed system) results in  the  appearance  of  some  unusual  but  commonly
accepted behavior for the system alone (open system) which would otherwise be
completely absent.  The most salient feature  of  the  dynamics  of  an  open
system$^{42}$ is dissipation, and the associated appearance of an
(thermodynamic)
arrow of time for the system observer.  Formally  one  sees  this  difference
arises  when  one  goes  from  the  density   matrix   description   of   the
system-environment obeying the unitary Louiville-von Neumann  equation,to  the
reduced density matrix for the system  alone  obeying  a  non-unitary  master
equation$^{43}$.  This is the basic reason for the apparent contradiction
between
the unitarity of microphysics and the  apparent  breakdown  of  macrophysics.
This schema can be used to describe  the  so-called  "collapse  of  the  wave
function",  or  quantum  to  classical  transition  in  quantum   measurement
theory$^{10,11}$.  The disappearance of interference between different
branches of
the  wave  functions  associated  with  the  vanishing  of  the  off-diagonal
components of the reduced density matrix is  called  decoherence.   It  is  a
consequence of information loss in the system through  its  coupling  to  the
environment.  In most practical cases, the environment can be simplified as a
stochastic source.  In fact, dissipative effects are oftentimes  depicted  by
phenomenological equations such as the  Langevin  equation.   Instead  of  an
explicit description of the environment, one replaces it with a noise  source
which drives the dynamics of the  system.   The  noise  source  can  be  from
thermal or vacuum fluctuations which are related to dissipation  through  the
fluctuation-dissipation relation.
\par
This general scheme in statistical mechanics, often modeled by the motion
of a Brownian particle (system) interacting with  a  collection  of  harmonic
oscillators (bath) and depicted by a master equation for the reduced  density
matrix$^{42-47}$ show  clearly  1)  the  interconnection  of  fluctuation,
noise,
dissipation, decoherence and correlation;  and  as  a  consequence  of  these
processes, 2) the nonspecificity of initial  conditions  in  determining  the
long time behavior of the system.
\par
We have applied this scheme to interpret quantum dissipative processes in
semiclassical gravity$^{29}$.  Let us now examine the issues of quantum
cosmology
in this light.  The issues  I  raised  earlier  can  be  grouped  into  three
distinct levels:
\par
\noindent (1) Is gravity an effective  theory?   If  yes,  how  do statistical
concepts enter?
\par
\noindent (2) Assuming that it is not, i.e., that the 3-geometries  can  be
viewed  as
fundamental physical variables, then within this (conventional) framework  of
quantum cosmology,
\par
a)  How valid is the minisuperspace approximation in describing the  full
dynamics of quantum cosmology?
\par
$b)$  How specific are the initial conditions in yielding classical limits?
\par
\noindent (3) What brings about decoherence and classicality?
\par
Note that questions in  level  (3)  are  actually  questions  in  quantum
mechanics proper, without a proper understanding of which one  cannot  get  a
satisfactory answer to the corresponding questions in quantum  cosmology.   A
thorough understanding of (2) alone may not help answering questions in level
(1), but there is reason to believe that answers to questions  in  level  (1)
could help to resolve most questions on level (2).  Cracking the  mystery  of
(1) is, however, very difficult.
\par
I will give here a general description of how statistical  considerations
enter into quantum cosmology and leave  the  more  technical  discussions  on
methodology and sample calculations to the next section.
\par
As we mentioned early, cosmology is a study  of  spacetime  dynamics  and
structure based on the observed facts in our  universe.   From  the  observed
high degree of isotropy and homogeneity one constructs the  standard  model
$^{18}$
based on the Friedmann-Robertson-Walker universe,  which  is  a  fairly  good
depiction of the late history of the universe back to at least $10^{-43}$ sec.
from the  big  bang.   The  flatness  and  entropy  problems  prompted  one
into speculating that at some very early time (the grand unification time
${\char"7E} 10^{-35} \sec )$ the universe might have undergone a stage of
inflation$^{20}$.  This  ushered
in renewed interest of the de  Sitter  universe.   Mathematical  analysis  of
Einstein's equation indicates that near the singularity  the  universe  could
have  been  highly  anisotropic  and  inhomogeneous,  thus  pointing  to  the
relevance of the Bianchi models (spatially  homogeneous)  and  the  mixmaster
universe (Type IX rotation group of motion) in particular$^{19}$.  There is
no  a
priori reason why the inhomogeneous cosmologies are usually  ignored,  except
that  they  are  very  difficult  to  study,  and   for   the   belief   that
inhomogeneities  and  anisotropies  in  the  very  early  universe  could  be
dissipated away through various mechanisms, vacuum particle  production  near
the Planck time being the most powerful$^{24,25}$.
\par
So cosmology is the study of the dynamics of a  very  restricted  set  of
spacetime (with high symmetries) to begin with.  This is similar to the first
task of statistical mechanics, i.e.,  selecting  a  few  relevant  parameters
which can capture  the  physical  essence  of  the  whole  system.   In  late
cosmology, these parameters are given by observation (e.g. Hubble  expansion,
microwave background), but in early cosmology especially in the  realm  where
quantum phenomena are dominant, which parameters  are  important  is  not  as
straightforward.  (It may be that spacetime is an averaged, composite concept
and even the simple scale factor in the metric  function  loses  its  obvious
meaning.  See  ideas  from  Regge  calculus$^{48}$,  random  geometry$^{39}$,
cellular automata$^{49}$, causal sets$^{50}$, etc).   Within  the
conventional  framework,  the
minisuperspace approximation in quantum cosmology assumes that  the  infinite
degrees of freedom ignored has  little  effect  on  the  selected  ones  (see
however calculation of Halliwell and Hawking$^{7})$.  Even so, using the
schema in statistical  mechanics  I  described  above,  this  is  highly
questionable. Viewing the scale factor a and the anisotropy parameters $\beta
\pm $  as  our  system, say, in a Bianchi Type IX universe$^{19}$  and
treating  the  remaining  infinite degrees  of  freedom  corresponding  to
other   anisotropic $(\theta _{1,2,3})$   and
inhomogeneous universes as the environment,  one would get  additional  terms
in the (non-unitary)  master  equation  associated  with  the  Wheeler-DeWitt
equation (an energy constraint condition$)^{31}$.  With an  eye  towards
matching the classical limit, one  can  use  the  Wigner  functional
$f(g,\pi )$  for  the 3-geometries $g$ and the conjugate momenta $\pi
($which describes the  distribution of  states  in  a  superphase  space)
and  obtain  a  Wheeler-DeWitt  Vlasov equation$^{28}$, or a Fokker-Planck
equation$^{11,22,19}$.  These equations embody  the
dissipative and diffusive effects  in  the  dynamics  of  the  minisuperspace
variables.  An immediate consequence is that the late time behavior  will  no
longer depend sensitively on the specific choice of initial conditions.   One
can show that for reasonable conditions on the bath and the coupling,  memory
loss (near Markovian behavior) is  a  rather  general  phenomena$^{30-33}$.
 This viewpoint supports the chaotic cosmology philosophy,  which  the
mixmaster$^{19}$ and the inflationary$^{20}$ programs both share - i.e.,
that the present state  of
the universe depends only weakly on the stipulation  of  initial  conditions,
but results largely from its own dynamics and interactions.
\par
For cosmology, splitting the system and the bath may pose some conceptual
difficulty if  one  envisages  the  Universe  as  containing  "everything"  -
spacetime and matter.  Though this is by definition true, it  is  in  reality
often not so stringently applied.  The observable "Universe" often refers  to
the causally connected parts which are much  smaller,  and  relevant  physics
goes on largely within the particle horizons.  One can take the  part  beyond
as the environment.  For spacetimes with event horizons (e.g. the  de  Sitter
universe)  physics  within  and  without  can  be  quite   different   (e.g.,
Starobinsky's  stochastic  inflation$)^{21}$.   One  can  also  view
topological fluctuations  such  as  wormholes  and  baby  universes  as
making  up   the environment, which can have interesting effects  on  the
universe  proper$^{51}$.
For an earlier discussion on how nontrivial global  structures  of  spacetime
and quantum processes in the universe can lead to entropy generation see Ref.
52.
\par
Another way statistical considerations enter into cosmology is from chaos
and dynamical systems.  As is well-known, the  Einstein  equations  for  some
classical  cosmological  models  (e.g.  Bianchi  Type   IX)   admit   chaotic
behavior$^{53}$.  The criteria of chaos in this context (e.g.  Liapunov
exponent) are still under investigation$^{54}$, but the existence of even
slight chaos  will
render  the  specificity  of  initial  conditions  highly  unstable  in   its
prediction of late time behavior.  Whether quantum dynamics exhibits the same
degree of chaos is unknown but it is unlikely  that  the  initial  conditions
could remain highly regulative and predictive,  as  the  prevailing  view  in
quantum cosmology seeks to establish.
\par
A related criticism of a statistical nature of  the  "initial  condition"
school  of   thought   comes   from   complexity   and   information   theory
considerations.  C. H. Woo$^{55}$ pointed out that in addition  to
a  simple  and
elegant initial  condition  for  the  wave  function  of  the  universe,  the
macroscopic  variables  require  arbitrary  inputs  for   a   more   detailed
description of the classical history of the universe.  He estimated that "the
number of bits needed to encode the algorithm for the wave function  and  its
simple initial condition is relatively small, of the order  of $10^{3}$".
Thus,
"such a wave function can predict at most the  specific  behavior  of  a  few
hundred macroscopic variables", a far cry from the sweeping optimism  implied
by the initial conditions school of thought.
\par
So far we have assumed that gravity is viewed as a fundamental field.  If
instead gravity is regarded as a composite force or an effective theory, then
statistical considerations would enter in an even  more  basic  and  familiar
way.  Think about how we construct molecules from atoms, then gas, fluids and
solids.  Statistical mechanics is an almost indispensable tool when one wants
to extract meaning  and  structure  from  a  collection  of  more  elementary
constituents.  In addition to the  effective  theory  of  Sakharov  mentioned
above, I have also toyed with the idea of viewing gravity as the result of  a
"time-dependent"  Hartree-Fock$^{56}$  interaction  among  the  unspecified
basic
constituents, similar to the nuclear  collective  model,  where  the  "normal
modes" of nuclear rotation  and  vibration  are  regarded  as  the  dynamical
variables.  Note that for the description of many gross features one does not
need to know the details of the  nucleons,  which  are  the  more  elementary
constituents, but only their collective motion.  This is in contradistinction
to the independent particle model, where the starting point is the individual
nucleon.  From this maybe we can learn something about  the  relationship  of
the two approaches to  quantum  gravity,  vis,  starting  with  a  collective
structure like geometry or starting with the  more  basic  constituents  like
superstrings or pre-geometry$^{18}$.  The other idea I have toyed with
somewhat is
to view Einstein's theory as the hydrodynamic (long wavelength)  limit  of  a
micro-theory of gravity.  There is of course no implied  "collision"  of  the
basic constituents in the corresponding gravitational "kinetic theory"  other
than their  nonlinear  interaction.   But  the  reduction  of  the  unitarity
dynamics of a microsystem to a macrosystem  exhibiting  irreversibility  like
the Boltzmann equation is an interesting analogous scheme.  The incorporation
of  hydrodynamic  fluctuations  and  phase  transition   ideas$^{57}$   into
the
consideration of transition from quantum to classical gravity may bring forth
some  new  insight.   In  the  more  main  stream  recent  developments,  how
statistical mechanics enters into random geometry, conformal field theory and
superstring as ways to relate to quantum gravity should  be  quite  apparent.
It is not my intention to delve into these areas here but I hope at  least  I
have convinced you that  a  good  knowledge  of  non-equilibrium  statistical
mechanics (field theory) is essential for understanding the basic  issues  of
quantum cosmology.
\par
Let me now start over from the  beginning  and  discuss  some  techniques
which we find useful to treat problems of this nature.
\section{Formalisms and Sample Problems}

As I mentioned in the beginning, because there exist many subtleties  and
conceptual pitfalls in quantum cosmology and our current understanding of the
statistical meaning of quantum mechanics is still vague, it is  perhaps  more
fruitful to begin the investigation on a  more  familiar  and  firmer  ground
before putting them to test in the volatile  setting  of  quantum  cosmology.
These phenomena are:  fluctuation,  noise,  dissipation,  particle  creation,
decoherence and correlation; the processes are:  separation (definition of
open system), coarse-graining, backreaction semiclassical limit, and quantum to
classical transition.   Various  authors  have  tried  to  attack  individual
phenomenon without due consideration of the interrelation with others -  e.g.
decoherence without taking into account  dissipation$^{58}$,  semiclassical
limit
without taking into account the nonlinearity of fields and nonseparability of
the background$^{59}$.  And because they want to do these  in  the  complexity
  of
quantum cosmology and to see the result in one strike, they have to make many
simplifying assumptions which  may  actually  wash  away  the  many  features
special to quantum cosmology  (e.g.  the  Born-Oppenheimer  separability  for
geometry and matter,  the  WKB  approximation  for  the  wave  function,  the
linearity of gravitational normal modes).
\par
The platform we choose to work on in this first stage of investigation is
quantum field theory, first in flat space  and  then  in  curved  space,  the
latter being the semiclassical limit of  quantum  gravity.     We  have  some
previous knowledge of how dynamical excitation of the vacuum in the  form  of
particle creation can act as a dissipative force in changing the dynamics  of
spacetimes$^{24,25}$.  We know how the closed-time-path formalism can yield a
real and causal equation of motion  for  the  geometry$^{25}$  and  provide a
correct statistical interpretation of particle creation as a  dissipative
process$^{29}$. We  recognize  the  effective  action  as  a  suitable
object  for  studying backreaction effects$^{25}$.  We also gained some
experience in the properties  of Wigner function as a classical distribution
function$^{26}$, and in  dealing  with
near-uniform  kinetic  systems.   These  were  the  ground  posts   we   have
established to explore the statistical properties of quantum fields  with  an
eye on the cosmological applications.
\par
The one big missing piece in this mosaic is noise  and  fluctuation.   We
believe there must  exist  some  fluctuation-dissipation  relation  even  for
nonequilibrium systems (not just for close  to  equilibrium  linear  response
systems), as suggested in the particle creation and backreaction  problems
$^{29}$. Even though the effect of  noise,  fluctuation  and  dissipation  are
rarely mentioned in quantum cosmology$^{60}$, we believe they should play as
essential  a
role as decoherence and  correlation,  as  they  are  intrinsically  related,
although they address different aspects of the basic issues.
\par
Thus the first task we set for ourselves in this program  is to formulate
a stochastic theory of quantum fields from first principle.  The criterion is
that it  should  contain  the  interconnection  of  all  the  above-mentioned
processes and can address all statistical properties  of  quantum  fields  in
curved spacetime.   The  two  major  useful  ingredients  we  found  are  the
influence functional and the coarse-grained effective action,  which  I  will
now briefly describe.
\par
These ideas are of course not new.  The former was established by Feynman
and Vernon$^{45}$ , the latter is a recasting  of  the  Zwanzig-Mori$^{43}$
projection
operator formalism  in  effective  action  forms.  The  influence  functional
captures the overall averaged effect of the environment on the  system,  from
which one can obtain the quantum master  equation  for  the  reduced  density
matrix.  This formalism was lately popularized by Caldeira and  Leggett$^{41}$
in their work on dissipative tunneling.  It has also  seen  application  to
the
inflationary cosmology$^{61}$.  We have worked with the closed-time-path
integral formalism of Schwinger and Keldysh$^{46}$ for particle creation  and
backreaction problems$^{25}$ and we know that  it  is  suitable  for
treating  non-equilibrium  quantum fields$^{26}$,  but  did  not  realize
that  it  is  just  the  influence functional formalism of Feynman-Vernon (in
fact  Schwinger's  paper  was  on Brownian motion$)^{62}$.  One can interpret
the $x$ and $x'$  paths as one  propagating
forward and the other backwards in time. After we discovered this  connection
all of our previous results on CTP formalism can be carried over easily.  The
other pivotal point is the incorporation of noise and stochastic  sources  in
purely quantum field theory terms without the presence  of  a  thermal  bath.
This lifts the restriction of near-equilibrium conditions customarily imposed
in the treatment (e.g. linear response  theory)  of  noise,  fluctuation  and
dissipation.  Indeed one major result we obtained from  this  program  is  to
show   that   a   general   fluctuation-dissipation   relation   exists   for
non-equilibrium quantum systems.  We have used this to improve  on  a  recent
result of Unruh and Zurek, as well as making new predictions.
\par
Our program was executed in successive stages$^{14-16}$,   starting  with  the
quantum mechanical problem of a Brownian particle bilinearly (Cxq) coupled to
a collection of harmonic oscillators as  bath,  generalizing  the  result  of
Caldeira-Leggett and Unruh-Zurek to non-local dissipation and colored  noise.
(We call the dissipation local and the noise white  if  the  dissipation  and
noise kernels are both delta functions.)  We obtain (for the first  time,  we
believe) an exact quantum master  equation  for  these  more  general  cases.
Contrary to what  we  originally  anticipated,  this  equation  describing  a
non-Markoffian  process  turns  out  to  be  not   so   complicated   as   an
integro-differential equation, but only  an  ordinary  differential  equation
with complicated time dependent coefficients, and is exact (see Eq. 7 of Ref.
63).  When we apply the Wigner transform to this quantum master equation,  we
obtain  for  the  Wigner  distribution  function  the  quantum  Fokker-Planck
equation with time-dependent diffusion coefficients.  The second  problem  we
looked at was a biquadratic $(\lambda x^{2}q^{2})$ coupling  between  the
system  and  bath variables, still in the quantum mechanical context.
Here we  carried  out  a perturbation analysis (in preparation for the
$\lambda \phi ^{4}$  theory)  up  to  quadratic order  in $\lambda $  and
derived  a  nonstationary  quantum  master  equation  with
nonlinearly generated dissipation and colored noise.  The third  problem  was
to do everything in field theory, first in flat space then in de Sitter space
making use of its conformally flat property, thus deriving for  the  Brownian
field with  nonlinear  nonlocal  dissipation  and  colored  noise  a  quantum
functional master equation and the associated Fokker-Planck equation for  the
Wigner distribution functional.
\par
This completes the first  stage  of  our  program.   We  have  used  this
equation  to  study  the  loss  of  coherence  in  some  interesting  quantum
mechanical problems and cosmological models$^{63}$.  We are now in the process
of
applying this formalism to  problems  in  semiclassical  gravity  (connecting
dissipation due to particle creation to noise  and  fluctuation),  stochastic
inflation (noise generation) and quantum  cosmology  (may  need  a  different
framework). In this path-integral framework one recovers the nice results  of
Unruh and Zurek$^{10}$ on decoherence$^{58}$,  and  can  also  see  its
relation  with dissipation  and  other  processes.   One  can  also  relate
the  degree  of decoherence with correlation (between the coordinate and
momenta  variables) as one set of criteria for classicality$^{64}$.  We
have  indicated  before  that particle creation in quantum fields can  bring
about  both  dissipation  and  decoherence.  Calzetta and Mazitteli$^{65}$ have
recently shown in the context  of quantum field  theory  in  curved  spacetime
that  particle  creation  is  a necessary and sufficient condition for
decoherence.  For stochastic inflation we have some doubts on the validity
of Starobinsky's scheme$^{21}$  in  generating
white noise for a linear field via dynamic truncation at the  event  horizon.
Instead, with  nonlinear  mode  coupling,  one  can  generate  noise  without
assuming a dynamic  truncation.   We  applied  the  coarse-grained  effective
action method$^{66}$ to calculate the effect of bath (high frequency modes)
on the
system (low-frequency modes) which are nonlinearly coupled.  Indeed,  colored
noise is generated with nonlocal dissipation in this problem, which  is  what
our general program predicts$^{16}$.
\par
The problem  of  backreaction  and  of  semiclassical  approximation  was
brought up recently in the context of  quantum  cosmology$^{59}$.
We  feel  that
without incorporating dynamical fluctuations  both  in  the  fields  and  the
geometry, and dealing with the  nonlinearity  and  nonadiabaticity  condition
squarely one cannot bring forth too much new beyond what we already know from
quantum  field  theory  in  curved  spacetime.    Quantum   gravity   is   an
intrinsically nonlinear theory.  Backreaction is only an approximate  concept
- it is meaningful only if one can separate some  background  geometry  apart
from the remaining (field or gravitational) degrees of freedom which is  only
possible in linearized theory and at energies lower than the  Planck  energy.
For quantum gravity at the Planck energy where full nonlinearity is at  work,
this separation is not easily attainable.  There is of course still  particle
creation - not only of quantum fields, but also gravitons - but  the  way  to
treat these quantum processes is not by the background field method to  which
quantum field theory in curved  spacetime  belongs,  but  rather  by  dealing
directly with the nonlinear interaction of fluctuations of both  gravity  and
fields  and  their   dynamical  excitations  (which  give  rise  to  particle
creation).  Our formulation can shed some light on the  nonlinearity  aspects
if one models the  gravity-field  interaction  in  the  form  of  field-field
interaction, although factorizability assumption for the density matrix of a
closed system in the influence functional method is still a  limitation  (see
however, Grabert {\sl et al} in Ref. 47).
\par
We have not yet embarked on a  full  scale  investigation  of  the  basic
issues in quantum cosmology using the framework we constructed, as the  issue
of time still poses a nontrivial problem for the path-integral formalism (see
Ref. 13 and Kiefer in Ref. ll).  However we have earlier done a  few  smaller
pilot problems to show 1) how dissipation in quantum cosmology can efface the
memory of initial conditions  and  2)  how  the  higher  gravitational  modes
usually discarded can introduce an effective dissipative term in the equation
of motion for the  lower  (minisuperspace)  modes.   The  first  problem  was
illustrated by Calzetta$^{30}$ with the example of a  linear  scalar  field
in  a
Robertson-Walker universe.  The reduced density matrix obeys an  equation  of
motion with a viscosity term containing a  nonlocal  kernel,  signifying  the
existence of non-Markovian dissipation  processes.   In  the  second  problem
Sinha$^{67}$ used a $\phi ^{4}$ scalar field in a Robertson-Walker universe
to  mimic  the nonlinear gravitational interaction.  She viewed the scale
factor $a$  and  the lowest (conformal) scalar field mode $\chi _{0}$ as the
background  and  studied  the coupling of the higher scalar  field  modes
$\chi _{n}$  with  themselves  and  their
backreaction on the background modes via the coarse-grained effective  action
method.  Backreaction shows up as a  dissipative  term  in  the  equation  of
motion for the lowest modes, in addition to the usual renormalized mass and a
shifted natural frequency.  These results exemplify our  earlier  claim  that
statistical effects  can  alter  ones  view  on  the  importance  of  initial
conditions and the validity of the minisuperspace truncation.  We still  need
to understand better how the  dissipative  effect  manifests  itself  in  the
dynamics described by different  "times" chosen  in  quantum  cosmology.   We
would also like to see how the "thermodynamic" arrow of time emerges from the
dissipation and decoherence processes in a quantum to  classical  transition.
These are part of our future work.

\section{Summary}

I have arranged the issues, processes and methodologies discussed in this
talk into a Table below.  Also noted therein are  some  related  problems  in
gravity and cosmology where  statistical  mechanical  considerations  can  be
fruitful.  To conclude we think more attention need be paid to the  following
aspects:
\par
\noindent 1)  Relevance of statistical considerations in quantum cosmology.
\par
\noindent 2)  Their role in addressing basic issues in theoretical physics.
\par
\noindent 3)  Interconnectedness of statistical processes such as
\par
a)  decoherence, dissipation and correlation
\par
b) noise, fluctuation and
\par
c)  particle creation, backreaction and semiclassical approximation.
\par
\noindent For example, quantum to classical transition involves all processes
in $a)$ and requires considerations of $c)$, but $b)$ also affects $a)$ and
engenders $c)$.
\par
Inquires on the statistical properties of the vacuum (noise, fluctuation,
excitation by dynamics, constraint by event horizon) in curved spacetime  and
quantum gravity can also shed light on the interconnection of basic issues in
quantum mechanics, general relativity and statistic mechanics (as manifested
in the Hawking effect$)^{68}$.
\par
\newpage
\noindent {\sl Issues:}
\par
\medskip
\noindent 1)  How large  is  the  class  of  initial  conditions which can
admit
classical spacetimes as solutions?  How regulative and predictive are
the specific initial conditions?
\par
\noindent 2)  How valid is the minisuperspace approximation?
\par
\noindent 3)  How does time emerge?  Is classical spacetime a  necessary
condition
for rendering time as we perceive it?
\par
\noindent 4)  Quantum to classical transition - criteria for classicality.
\par
\noindent 5)  Semiclassical limit - relation of  quantum  field  theory  in
curved
spacetime with quantum cosmology.
\par
\noindent 6)  Separability of background  and  field,  validity  of  the
adiabatic
condition, backreaction and consistency.
\par
\medskip
\noindent {\sl Processes}
\par
\medskip
\noindent 1)  Coarse-graining:   how  sensitive are the final results to   the
averaging measure.
\par
\noindent 2)  Decoherence, correlation and dissipation
\par
\noindent 3)  Noise, fluctuation and dissipation
\par
\noindent 4)  Particle creation
\par
\medskip
\noindent {\sl Frameworks}
\par
\medskip
\noindent 1)  Nonunitary  evolution  equations: master equation,
Fokker-Planck
equation and Langevin equation.
\par
\noindent 2)  Closed-time-path integral formalism, influence-functional
formalism
\par
\noindent 3)  Superscattering (\$) matrix formalism$^{69}$.
\par
\medskip
\noindent {\sl Techniques}
\par
\medskip
\noindent 1)  Subdynamics and  projection  operator formalism;   coarse-grained
effective action (for coarse-graining).
\par
\noindent 2)  Wigner distribution function, coherent state representation  (for
classical limits).
\par
\noindent 3)  BBGKY hierarchy, nth order correlation function (for
correlation).
\par
\medskip
\noindent {\sl Other Related Problems}
\par
\medskip
\noindent 1)  Gravitational  entropy$^{70}$  and  Hawking effect: a  stochastic
field theoretical interpretation.
\par
\noindent 2)  Tunneling, decoherence and dissipation.
\par
\noindent 3)  Dynamical critical phenomena and noise-induced transition.
\par
\medskip
\noindent{\bf Acknowledgement}
\par
The work I described in this talk was done jointly in stages with Esteban
Calzetta, Juan Pablo Paz, Sukanya Sinha, and Yuhong Zhang with  whom  I  have
enjoyed many  interesting  discussions  and  correspondences.   (They  should
however not be held responsible for any outlandish comment or  crazy  idea  I
advanced here.)  I would like to  thank  the  organizers  of  this  workshop,
Professor Arimatsu in particular, for the effort they put in,  and  the  warm
hospitality they extended to us.  This work  is  supported  in  part  by  the
National Science Foundation under grant No. PHY-8717155

\newpage
\centerline{\bf  REFERENCES}
\par
\medskip
\noindent 1)  A. Ashtekar and J. Stachel (ed.) Conceptual Problems in Quantum
Gravity, Proceedings Second Osgood Hill Conference (Birkha\"user, Boston,
1989).
\par
\medskip
\noindent 2)  S. Coleman, J. Hartle, T. Piran and S. Weinberg (ed.)  Quantum
Cosmology and  Baby  Universes,  Proceedings  7th  Jerusalem  Winter School for
Theoretical Physics (World Scientific, Singapore, 1990).
\par
\medskip
\noindent 3)  W. H. Zurek (ed.) Complexity, Entropy and the Physics of
Information,
Proceedings Santa $Fe$ Institute Studies in  the  Sciences  of  Complexity,
Vol. IX. (Addison-Wesley, Reading, 1990).
\par
\medskip
\noindent  4)  B. S. DeWitt, Phys. Rev. {\sl 160} (1967) 1113.
\par
\medskip
\noindent 5)  J. A. Wheeler, in: Battelle Recontres, ed. C. DeWitt and
J.  A.  Wheeler (Benjamin, New York, 1968).
\par
\medskip
\noindent 6)  C. W. Misner, in: Magic Without  Magic,  ed.  J.  Klauder
(Freeman,  San Francisco, 1972).
\par
\medskip
\noindent 7)  J. B. Hartle and S. W.  Hawking,  Phys.  Rev. ${\sl D28} (1983)
2960$;   J.  J. Halliwell and S. W. Hawking, ibid ${\sl D31} (1985) 1777.$
\par
\medskip
\noindent 8)  A. Vilenkin, Phys. Lett. ${\sl ll7B} (1982) 25,$ Phys. Rev.
${\sl D27} (1983) 2848; {\sl D30} (1984) 509.$
\par
\medskip
\noindent 9)  J. J. Halliwell, Bibliography on Quantum Cosmology, Int. J. Mod.
Phys. (1990).
\par
\medskip
\noindent 10) J. A. Wheeler and W.  H.  Zurek  (ed.)  Quantum  Theory  and
Measurement
(Princeton Univ. Press, Princeton, 1983); W. H.  Zurek,  Phys.  Rev. ${\sl D24}
(1981) 1516; {\sl D26} (1982) 1862$; W. G. Unruh and W. H. Zurek, Phys.  Rev.
{\sl D40} (1989) 1071.
\par
\medskip
\noindent  11) H. D. Zeh, Found. Phys. {\sl 1} (1970) 69; Phys. Lett. {\sl
A116}
 (1986) 9.
E. Joos and H. D. Zeh, Z. Phys. ${\sl B59} (1985) 223$; E. Joos, Phys.  Rev.
${\sl D36} (1987) 3285.$
C. Kiefer, Class. Quant. Grav. {\sl 4} (1987) 1369; {\sl 6} (1989) 651;  Phys.
  Lett. {\sl A139} (1989) 201; "Interpretation of the Decoherence Functional
in  Quantum Cosmology" Zurich Preprint 1990.
\par
\medskip
\noindent 12) R. Griffith, J. Stat. Phys. {\sl 36} (1984) 219; R. Omn\`es,
ibid {\sl 53}  (1988)  893, 933, 957.
\par
\medskip
\noindent  13) M. Gell-Mann and J. B. Hartle in Ref. 3; J. B. Hartle in Ref. 2.
\par
\medskip
\noindent  14) Yuhong Zhang, Ph.D. Thesis, University of Maryland 1990.
\par
\medskip
\noindent 15) B. L. Hu, J. P. Paz and Y. Zhang, "Quantum Brownian Motion with
Nonlocal Dissipation and Colored Noise", Paper I.
\par
\medskip
\noindent 16) B. L. Hu, J. P. Paz and Y. Zhang, "Stochastic Properties  of
Interacting Quantum Fields", Paper II.
\par
\medskip
\noindent 17) B. L. Hu, J. P. Paz and Y. Zhang, "Noise and  Decoherence  in
Stochastic Inflation", Paper III.
\par
\medskip
\noindent 18) See,  e.g.  J.  Peebles,  Physical   Cosmology   (Princeton
University, Princeton, 1971) S. Weinberg, Gravitation and Cosmology (Wiley,
New York, 1982);
C. W. Misner, K. S. Thorne and Wheeler, Gravitation (Freeman,  San
Francisco, 1973).
\par
\medskip
\noindent 19) C. W. Misner, Phys. Rev. Lett. {\sl 22} (1969) 1071;  V.  A.
Belinsky,  I.  M. Khalatnikov and E. M. Lifshitz, Adv. Phys. {\sl 19} (1970)
525, {\sl 31} (1982)  639;
D. M. Eardley, E. P. T. Liang and R. K. Sachs, J. Math. Phys.  {\sl 13}  (1972)
99; M. P. Ryan and L.  C.  Shepley,  Homogeneous  Relativistic  Cosmology
(Princeton University, Princeton, 1975).
\par
\medskip
\noindent 20) A. Guth, Phys. Rev. ${\sl D23} (1981) 347$; A. Albrecht
and P.  J. Steinhardt, Phys. Rev. Lett {\sl 48} (1982) 1220;
A. Linde, Phys. Lett. ${\sl 108B} (1982) 389.$
\par
\medskip
\noindent 21) A. A. Starobinsky, in: Field Theory, Quantum Gravity and Strings,
eds. H. J. de Vega and N. Sanchez (Springer, Berlin,  1986);
S.  J.  Rey,  Nucl. Phys. ${\sl B284} (1987) 706$;
J. M. Bardeen and G. J. Bublik, Class. Quan. Grav. {\sl 4} (1987) 573;
F. Graziani, Phys. Rev. $D38 (1988) 1122, 1131, 1802.$
\par
\medskip
\noindent 22) A. A. Starobinsky, in: Quantum Mechanics in  Curved  Spacetime,
eds.  J. Audretsch and V. de Sabbata (Plenum, London, 1990).
\par
\medskip
\noindent 23) See, e.g., N. Birrell and P. C. W. Davies, Quantum Fields in
Curved Space (Cambridge University, Cambridge, 1982); B. L. Hu, L. Parker and
D. J. Toms,  Gravitation,  Quantum  Fields  and  Curved  Spacetime  (Cambridge
University, Cambridge, 199?).
\par
\medskip
\noindent 24) Ya. B. Zel'dovich and A. A. Starobinsky, JETP {\sl 34} (1972)
1159;  B.  L. Hu and L. Parker, Phys. Rev. ${\sl D17} (1978) 933$;
J. B. Hartle and B. L. Hu, ibid ${\sl D20} (1979) 1757, 1772$;
J. Frieman, ibid ${\sl D39} (1989) 389.$
\par
\medskip
\noindent  25) E. Calzetta and B. L. Hu, Phys. Rev. ${\sl D35} (1987) 495.$
\par
\medskip
\noindent  26) E. Calzetta and B. L. Hu, Phys. Rev. ${\sl D37} (1988) 2878.$
\par
\medskip
\noindent  27) E. Calzetta and B. L. Hu, Phys. Rev. ${\sl D40} (1989) 656.$
\par
\medskip
\noindent  28) E. Calzetta and B. L. Hu, Phys. Rev. ${\sl D40} (1989) 380.$
\par
\medskip
\noindent  29) B. L. Hu, Physica {\sl A158} (1989) 399.
\par
\medskip
\noindent  30) E. Calzetta, Class. Quant. Grav. ${\sl 6} (1989) L227.$
\par
\medskip
\noindent 31) B. L. Hu, "Quantum and Statistical Effects in Superspace
Cosmology"  in:
Quantum Mechanics in Curved Spacetime, Proceedings of the 11th Course  of
the International School on Cosmology and Gravitation, Erice 1989, ed. by
J. Audretsch and V. de Sabbata (Plenum, London, 1990).
\par
\medskip
\noindent 32) E. Calzetta, "Anisotropy  Dissipation  in Quantum  Cosmology",
IAFE
preprint, Buenos Airs, 1990.
\par
\medskip
\noindent 33) E.  Calzetta  and  B.  L. Hu  "Dissipation  in  Quantum
Cosmology",  in preparation.
\par
\medskip
\noindent  34) J. P. Paz, Phys. Rev. ${\sl D40} (1990) 1054.$
\par
\medskip
\noindent  35) J. P. Paz, Phys. Rev. ${\sl D42} (1990) 2.$
\par
\medskip
\noindent 36) See, e.g., M. Green, C. Schwarz and E. Witten, Superstrings Vol.
$I \&$  II
(Cambridge University, Cambridge, 1987); M. Kaku, Superstrings (Springer,
Berlin,  1988);  W.  Siegel,  String  Field  Theory  (World   Scientific,
Singapore, 1989).
\par
\medskip
\noindent 37) See, e.g., B. E. Baaquie {\sl et al} (ed.) Conformal  Field
Theory,  Anomalies and Superstrings (World Scientific, Singapore, 1988).
\par
\medskip
\noindent  38) See, e.g., E. Witten, Comm. Math. Phys. {\sl 117} (1988) 353.
\par
\medskip
\noindent 39) See, e.g., D. Gross and A. A. Migdal, Phys. Rev. Lett. {\sl 64}
(1990) 127;  E. Brezin and V. Kozakov, Phys. Lett. ${\sl B236} (1990) 144$;
M. R. Douglas and  S. H. Shenkar, Nucl. Phys. ${\sl B335} (1990) 635$;
M. R. Douglas, Phys. Lett. ${\sl B238} (1990) 176.$
\par
\medskip
\noindent 40) A. D. Sakharov, Dok. Akad. Nauk. SSSR {\sl 177} (1967) 70
[Sov. Phys.- Doklady  {\sl 12}
(1968) 1040]; S. L. Adler, Phys. Rev. Lett. {\sl 44}  (1980)  1567;  Rev.  Mod.
Phys. {\sl 54} (1982) 719; A. Zee, Phys. Rev. Lett. {\sl 42} (1979) 417.
\par
\medskip
\noindent 41) K. Kuchar and M. P. Ryan, Jr., in: Proc. Yamada Conference XIV.
eds.  H. Sato and T. Nakamura (World Scientific, Singapore, 1986),
Phys. Rev. ${\sl D40} (1989) 3982.$
\par
\medskip
\noindent  42) E. B. Davies, Quantum Theory of Open Systems (Academic, London,
1976).
\par
\medskip
\noindent 43) R. Zwangzig in: Lectures in Theoretical Physics III, eds. W. E.
Britten, B. W. Downes and J. Downes (Interscience, New  York,  1961)  106-141;
H. Mori, Prog. Theor. Phys. {\sl 33} (1965) 1338;
C. R. Willis and R.  H.  Picard, Phys. Rev. {\sl A9} (1974) 1343;
R. Balescu,  Equilibrium  and  Nonequilibrium,
Statistical Mechanics (Wiley, New York,  1975);  H.  Grabert,  Projection
Operator   Techniques    in    Nonequilibrium    Statistical    Mechanics
(Springer-Verlag, Berlin, 1982);  B. L. Hu and H. E. Kandrup, Phys.  Rev.
${\sl D36} (1987) 1776$ and references therein.
\par
\medskip
\noindent 44) R. Rubin, J. Math Phys. {\sl 1} (1960) 309, {\sl 2} (1961) 373;
G. W. Ford, M. Kac and P. Mazur, J. Math. Phys. {\sl 6} (1963) 504.
\par
\medskip
\noindent 45) R. P. Feynman and F. L. Vernon, Ann. Phys. (N.Y.) {\sl 24} (1963)
118;  R.  P. Feynman  and  A.  R.  Hibbs,  Quantum  Mechanics and Path
Integrals (McGraw-Hill, New York, 1965).
\par
\medskip
\noindent 46) J. Schwinger, J. Math. Phys. {\sl 2} (1961) 407; L. V. Keldysh,
Zh. Eksp. Teor. Fiz. {\sl 47} (1964) 1515 [Sov. Phys. JETP 20 (1965) 1018].
K. C. Chou, Z.  B. Su, B. L. Hao and L. Yu, Phys. Rep. 118 (1985) 1.
\par
\medskip
\noindent  47) A. O. Caldeira and A. J. Leggett, Physica (Utrecht) $121A (1983)
 587$;
H. Grabert, P. Schramm and G. Ingold, Phys. Rep. {\sl 168} (1988) 115.
\par
\medskip
\noindent 48) T. Regge, Nuovo Cimento 19 (1961) 558; R. Sorkin, Phys. Rev.
${\sl D12}
(1975) 385$; H. Hamber and R. M. Williams, Nucl.  Phys. ${\sl B248} (1984)
392, {\sl B267} (1986) 482, {\sl B269} (1986) 712$; R. Friedberg and T. D. Lee,
Nucl. Phys. ${\sl B242} (1984) 145$; J. B. Hartle, J. Math. Phys. {\sl 26}
(1985) 804, {\sl 27} (1986) 287. \par
\medskip
\noindent 49) An  example  of  ideas  in  cellular  automata  applied  to
inflationary cosmology  is  Linde's  eternal,  self-reproducing  universe:
A.  Linde, Physica Scripta ${\sl T15} (1987) 169.$
\par
\medskip
\noindent 50) See, e.g., L. Bombelli, J. Lee, D. Mayer and R. D. Sorkin, Phys.
Rev. Lett. {\sl 59} (1988) 521.
\par
\medskip
\noindent  51) J. Ellis, S. Mohanty and D. V. Nanopoulos, Phys. Lett.
${\sl B211} (1989) 113.$
\par
\medskip
\noindent 52) B. L. Hu, in: Cosmology of the Early Universe eds. H. Sato and
R. Ruffini (World Scientific, Singapore, 1984) and references therein.
\par
\medskip
\noindent 53) J. D. Barrow, Phys. Rep. {\sl 85} (1982) 1; I. M. Kalatnikov
{\sl et al}  in:  General
Relativity and Gravitation $(GR10,$ Padova, 1983) ed. B.  Bertotti,  F.  de
Felice and A. Pascolini (Reidel, Dordrecht, 1984); O. I.  Bogoiavlenskii,
Methods in the Qualitative Theory of Dynamical  Systems  in  Astrophysics
and Gas Dynamics (Springer, Berlin, 1985).
\par
\medskip
\noindent 54) A. Ashtekar and J. Pullin, Syracuse Preprint 1990; K. Schleich,
talk at Quantum Cosmology Workshop,  UBC,  May  1990;
E.  Calzetta,  "Cosmology, Entropy and Chaos" IAFE Preprint, Buenos Airs 1990.
\par
\medskip
\noindent 55) C. H. Woo, Phys. Rev. ${\sl D39} (1989) 3174, {\sl D41} (1990)
1355,$
and in Ref. 3; K. Svozil, Phys. Rev. ${\sl D41} (1990) 1353.$
\par
\medskip
\noindent 56) A. Bohr and B. R. Mottelson, Nuclear Structure,  Vol.  2
(Benjamin,  New York,  1975);  K.  Goeke  and  P.  -G.  Reinhard  (eds.)
Time-Dependent Hartree-Fock and Beyond (Lecture Notes in Physics 171)
(Springer, Berlin, 1982).
\par
\medskip
\noindent 57) D. Forster, Hydrodynamics Fluctuations, Broken Symmetry and
Correlation Functions (Benjamin, New York, 1975).
\par
\medskip
\noindent 58) See e.g., J. J. Halliwell, Phys. Rev. ${\sl D39} (1989) 2912$;
T. Padmanabhan ibid ${\sl D39} (1989) 2924$ and references therein.
\par
\medskip
\noindent 59) See e.g., T. P. Singh and T. Padmanabhan, Ann. Phys.  (N.Y.)
{\sl 196}  (1989) 296 and references therein.
\par
\medskip
\noindent 60) There are scattered work on this aspect  but  are  mostly
incomplete  or incorrect,  see,  e.g.,  M.  Morikawa,
Phys.  Rev. ${\sl D33} (1986) 3607,$ ibid ${\sl D40} (1989) 4023$;
ibid ${\sl D42} (1990) 1027.$
\par
\medskip
\noindent  61) J. M. Cornwall and R. Bruinsma, Phys. Rev. ${\sl D38} (1988)
3146.$
\par
\medskip
\noindent 62) In this workshop Professor Su kindly quoted me a paper he wrote
[Z. B. Su, L. Y. Chen, X. T. Yu, K. C. Chou, Phys. Rev. ${\sl B37} (1988)
9810]$
which shows the equivalence of these two formalisms.
\par
\medskip
\noindent  63) J. P. Paz,
Nonequilibrium Quantum Fields in Cosmology, this volume.
\par
\medskip
\noindent 64) J. J. Halliwell, Phys. Rev. ${\sl D36} (1987) 3626$;
H.  Kodama,  Prog.  Theor. Phys. (1989); S. Habib and R. Laflamme,
"Wigner Function and  Decoherence
in Quantum Cosmology", UBC  Preprint  (1990);  A.  Anderson,  Univ.  Utah
Preprint (1990); J. P. Paz and S. Sinha, "On Decoherence and Correlation"
in preparation.
\par
\medskip
\noindent 65) E. Calzetta and F. D. Mazzitelli, "On Decoherence and Particle
Creation" IAFE Preprint (1990).
\par
\medskip
\noindent 66) B. L. Hu and Y. Zhang, "Coarse Grained Effective Action,
Renormalization Group Transformation and Inflationary Cosmology"
University  of  Maryland Preprint (1990).
\par
\medskip
\noindent 67) B. L. Hu and S. Sinha, "Coarse-Grained Effective Action,
Backreaction and Minisuperspace Approximation" University of Maryland
preprint (1990).
\par
\medskip
\noindent  68) S. W. Hawking, Nature {\sl 248} (1974) 30.
\par
\medskip
\noindent 69) S. W. Hawking, Comm. Math. Phys. {\sl 87} (1982) 395; S.  W.
Hawking, Physica Scripta ${\sl T15} (1986)$; D. Page, Phys. Rev. ${\sl D34}
(1986) 2267.$
\par
\medskip
\noindent 70) R. Penrose, in: General Relativity, An Einstein Centenary (eds.)
S.  W. Hawking and W. Israel (Cambridge Univ., Cambridge, 1979);  B.  L. Hu,
Phys. Lett. ${\sl 97A} (1983) 368.$
\par
\end{document}